# Chemical instability of free-standing boron monolayers and properties of oxidized borophene sheets

Xue Lei,[1] Anatoly F. Zatsepin,[1] Danil W. Boukhvalov[*,2,1]

[1]Institute of Physics and Technology, Ural Federal University, Yekaterinburg 620002, Russia

[2]College of Science, Institute of Materials Physics and Chemistry, Nanjing Forestry University, Nanjing 210037, P. R. China

*In this work we report results of step-by-step modeling of the oxidation of free-standing boron monolayers of different types. Results of the calculations demonstrate that the process of the oxidation is always exothermic and lead toward the formation of foam-like boron oxide films with incorporated non-oxidized small boron clusters. Some of these boron-oxide films demonstrate the presence of chemically stable magnetic centers. Evaluation of the physical properties of oxidized boprophene sheets (OBS) demonstrate it possible application in solar energy, as sensors and coating against leakage of hydrogen.*

e-mail: danil@njfu.edu.cn

## 1. Introduction

Since the successful isolation of graphene, two-dimensional (2D) materials have been a focus of research in electronic devices due to their unique physical properties and potential applications. [1, 2, 3] Successful production of multiple new 2D compounds stimulates theoretical design of future members of this class. There two main approaches to the theoretical predicting of novel 2D materials. The first is consideration of the layers as naturally layered structures as separate monolayer (for example graphene and graphite, phosphorene and black phosphorous). The second approach is varying of the chemical composition of already synthesized 2D structures with keeping of geometrical structure (for example multiple layered dichalcogenides beyond $MoSe_2$). The third approach is theoretical design of completely new compounds (for example silicene and germanene). Recent synthesis of silicene [4] demonstrate efficacy of latter approach.

Boron is the neighbor of carbon in the periodic table and similarly to carbon has a multiple valence. These facts stimulate modeling of boron analogous of carbon low dimensional structures. Initially was discussed stability of boron-based materials with fullerene-like boron structures [5], boron nanotubes [6] and further 2D boron. [7] Dozens of simulated allotropes of 2D boron with similar atomic structure and almost the same total energy per boron atoms were reported. The nature of this multiplicity of 2D boron allotropes was also discussed. [8] Successful synthesis of silicene on metallic substrate [9] stimulate the modelling of the grown of 2D boron over the most used for the grown of 2D systems metallic substrates. [10].

Planar hexagonal B36 was proposed as potential bases for extended 2D borophene [11]. After that, four phases (2-Pmmn, β12, χ3 and graphene-like phases) of borophene have been synthesized on Ag (111) surface or Al (111) surface substrates under ultrahigh vacuum conditions [8, 12, 10]. The 2-Pmmn phase of borophene is a corrugated configuration with highly anisotropic electronic and mechanical properties, β12, χ3 are planar without vertical undulations, for graphene-like borophene on Al substrate is more energetically stable than on Ag substrate [13]. In 2007, Yakobson's group [14] predicted the existence of mechanically stable $B_{80}$ fullerene cage similar to $C_{60}$ buckyball. However, hollow cages are indeed not the ground state structures for the medium-sized BN clusters from N=68, whereas core-shell configurations are more thermodynamically preferable based on ab initio global search by different groups [15-18]. The energetic unfavorability of these empty boron cages can be attributed to the electron deficiency of boron, which tends to form more compact structures.

In recent works were reported that borophene is a highly anisotropic metal, the optical properties of borophene exhibit strong anisotropy as well [8, 19] and superconductivity induced by the strain. [20] Another experiments reports that the only line-edged borophene nanoribbons are stable in the free-standing form and demonstrate low-resistivity Ohmic conductance [21]. 2D borophene was proposed as an ideal electrode material with high electrochemical performance for both Li-ion and Na-ion batteries [22]. The stability of lattice defects including boron vacancy was investigated, they found find that the mechanical strength of borophene is reduced by the vacancy, the anisotropy in Poisson's ratio can be tuned [23]. Studied on the adsorption behavior of borophene towards various molecules by using density functional theory method [24-28]. Borophene was also discussed as prospective material for gas sensing. [27,28] For bilayer structures the AA stacking mode was found the most stable among the six high-symmetry stacking configurations. Since the layered structures can withstand larger critical strains than that in monolayer, layered borophene exhibits more flexibility than monolayer one [29]. $NiB_6$ monolayer was proposed as a stable 2D Dirac material with anisotropic elastic properties with a Young's modulus of 189 N m$^{-1}$. The calculation of band indicated that a double Dirac cone feature near the Fermi level with a high Fermi velocity of 8.5×10$^5$ m s$^{-1}$[30].

Stimulated by the $B_{80}$ fullerene cage, a 2D boron sheet (BS), namely "α-sheet", with appreciable stability was constructed [31-33]. By incorporating periodic hexagonal holes in the triangular lattice to reach the balance between three-center (3c) and two-center (2c) bonds, this mixed hexagonal-triangular boron sheet (ht-BS) is more stable than previously proposed buckled triangular boron sheets (t-BS) [34-37], which can be explained by a chemical bonding picture that the hexagon holes serve as scavengers of extra electrons from the filled hexagons [38]. In recent works were reported results of the seeking of more stable structure for 2D allotropes of boron, including the snub-sheet, g1/8-sheet, g2/15-sheet, α1-sheet, β1-sheet, struc-1/8-sheet [39-42]. All these ht-BSs can be constructed by carving different patterns of hexagonal holes within the triangular sheet and described by a hexagon hole density η, defined as the ratio of number of hexagon holes to the number of atoms in the original t-BS.

These works motivate experimentalist to further attempts to synthesize free-standing borophene. In previous experiments [43,44] single-walled and multi-walled boron nanotubes have been observed. With metal passivation to stabilize the sp2 hybridization, similar silicene monolayer has been successfully fabricated on Ag(111) and Ir(111) surfaces in recent experiments [44-48]. On the other hand, small boron clusters in vacuum were proved to adopt quasi-planar configurations [9,49-54], which may act as precursors for experimental synthesis of BS on metal surface via soft-landing of cluster beams [55].

Note that the synthesis of BS was done in the absence of oxygen but possible application in devices usually will be at ambient conditions. Fast degradation of the phosphorene [56,57] and oxidation of dichalcogenides [58] suggest that additionally to mechanical stability chemical stability should be also discussed. In our work we provide modelling of step-by-step oxidation of two representative types of borophene sheets and discuss the properties and possible applications of obtained oxidized boron sheets (OBS).

## 2. Computational Method

Modeling was performed using density functional theory (DFT), implemented by means of the pseudopotential code SIESTA. [59] All calculations were performed using the generalized gradient approximation (GGA-PBE) including spin polarization. [60] The calculations of physical adsorption were performed with taking into account van der Waals correction [61]. During this optimization, the ion cores were described by norm-conserving nonrelativistic pseudopotentials [62] with cutoff radii 1.59, 1.47, 1.48 and 1.25 a.u. B, O, N and H respectively. The wavefunctions were expanded with a double-ζ plus polarization basis of localized orbitals. These calculations were carried out with an energy mesh cut-off of 300 Ry and a k-point mesh of 4×4×2 within the Monkhorst-Pack scheme [63]. For the plot of the densities of states (DOS) the number of k-point were increased up to 8×8×2. The separation between the slabs within periodically boundary condition is 2 nm.

## 3. Results and discussions

### 3.1. Modeling of the oxidation process

Multiple structural phases of BS reported in the previous works. [39-42] These structural phases have almost the same total energy per boron atom. This multiple structural forms of BS can be classified in several groups denoted by Greek letters. The most energetically favorable is so-called β and χ types of BS. The first type can be described as continuous BS with triangular lattice with the pairs of hexagonal pores nearby (Fig. 1). Ne number of these pairs of hexagonal pores per unit of square can vary within this type. The second is the same triangular lattice with the lines of hexagonal pores (Fig. 2). The distance between the pores can be different for BS of this type. Two structures (one from each class of BS) with minimal number of the atoms in the supercells have been chosen for the further modeling.

To reveal the sites for oxidation by molecular oxygen we locate the oxygen atoms above the nearest boron atoms in different non-equivalent places of BSs and after optimization find the configuration with the lowest total energy. For both types of studied BSs we have find that the first step of the oxidation will be decomposition of molecular oxygen on the edges of hexagonal pore (Figs. 1,2). For the check of the favorability of this process we performed the calculation of formation energy by standard formula:

$E_{form} = E(host+guest) - [E(host) + E(guest)]$,

where E(host+guest)] is the energy of the system after decomposition of the oxygen atom, E(host) is the total energy of the system before this process and E(host) is the energy of oxygen molecule in ground (triplet) state in gaseous phase. Results of the calculations demonstrate that for both types of considered BSs this process is extremely energetically favorable. Because the magnitude of formation energy is several times larger than activation energy of the oxygen on pure and doped graphene [64] the process should be barrierless.

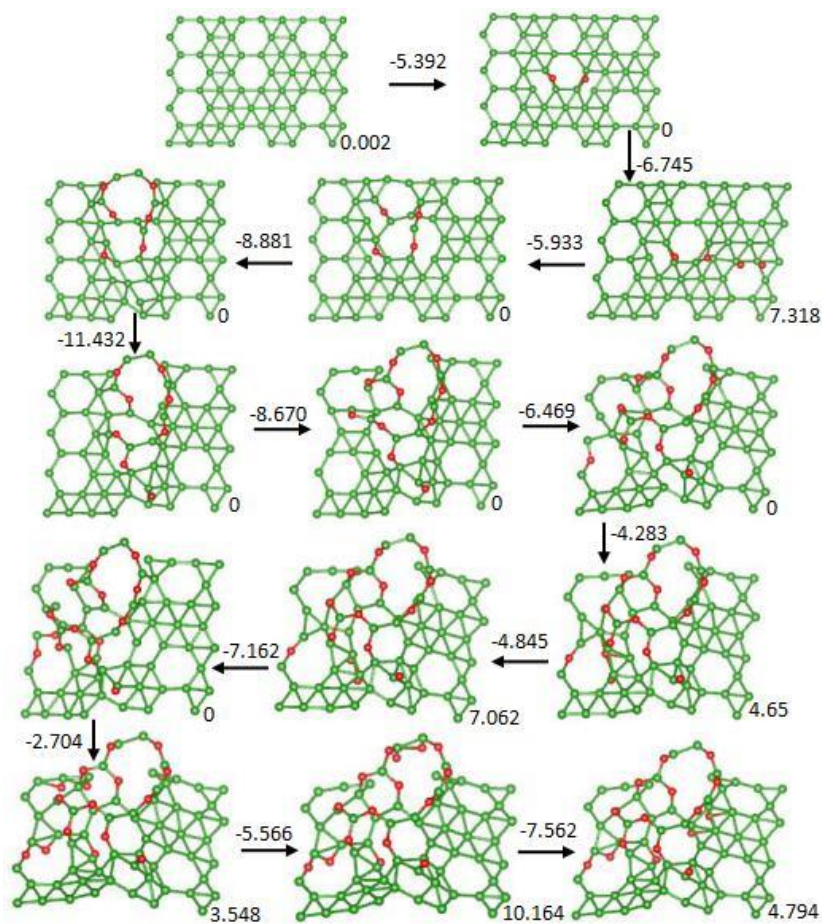

**Figure 1.** Optimized atomic structure of the first steps of the oxidation process of boron sheet of β-type. The numbers near the arrows is formation energies (in eV) and numbers in the corner of the structures is total magnetic moment of the supercell (in $\mu_B$).

Significant favorability of the first step of the oxidation suggest us to perform further modelling of step-by-step oxidation. Because distortion of the monolayer in result of the first step of the oxidation increase the number of non-equivalent pairs of atoms we considered only decomposition of the next oxygen molecules in vicinity of the hexagonal pores and on the boron atoms with the largest out-of-plane distortions. Results of the calculations (Figs. 1,2) evidence similarity of the patterns of further oxidation for both types of BSs: oxidation continue first on the borders of hexagonal pores and then goes deeper in the areas of continuous triangular lattice. All steps of the oxidation process of both types of BSs is also significantly favorable. This energetic favorability persists when oxidation process turns from edges of the hexagonal pores into the areas of triangular lattice. Further steps of the oxidation are also energetically favorable and lead formation of foam-like structures in both considered types of BSs (see Fig. 3). Note that in both studied structures in the process of oxidation occur formation non-oxidized boron cluster with atomic structure similar to $B_{12}$. [17] Based on results of the modelling that demonstrate similarity for two representative samples of different types of BSs can propose that our results are valid for all BSs.

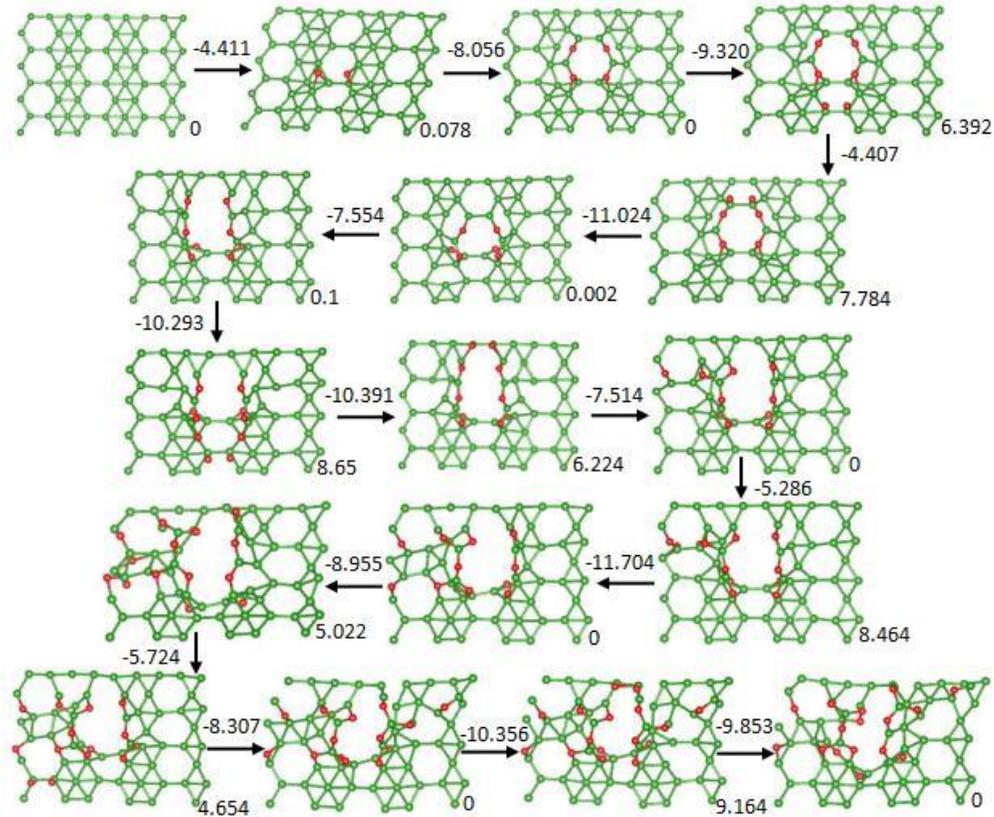

**Figure 2.** Optimized atomic structure of the first steps of the oxidation process of boron sheet of χ-type. The numbers near the arrows is formation energies (in eV) and numbers in the corner of the structures is total magnetic moment of the supercell (in $\mu_B$).

Therefore, similarly to the phosphorene, [56,57] BSs require fabrication of additional protective cover made from chemically inert material (for example hBN). [65] From the other hand, oxidized BSs (OBSs) can be interesting by itself and further we provide evaluation of the properties of these materials that can be used in further applications.

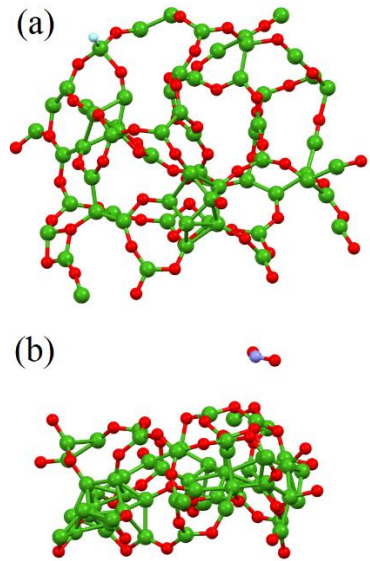

**Figure 3.** Optimized atomic structure of totally oxidized boron sheet of β-type with magnetic center passivated by hydrogen atom (a) and totally oxidized boron sheet of χ-type with physically adsorbed molecules of $NO_2$.

## 3.2. Magnetic properties and electronic structure

In contrast to graphene where formation of epoxy groups does not induce any magnetic moments oxidation of both considered types of BSs provides appearance of dangling bonds which is the source of unpaired electrons. [67] The cause of this difference is odd number of electrons in boron *2p* shell. At early step of oxidation magnetic moments is rather large because adsorption of the oxygen atoms provides break of multiple bonds. The values of magnetic moments per supercell at early stages of the oxidation is non-integer because BSs remain metallic. Further oxidation until final step provides saturation of all dangling bonds in χ-OBS but surprising survival of magnetism (1 $\mu_B$ per cell) in β-OBS. Usually oxidation provide vanishing of *sp*-magnetism (also called *d0*-magnetism) by saturation of all dangling bonds but in the case of β-OBS where special topology prevents passivation of magnetic center. [67] Formation of this magnetic state in β-OBS is related with appearance of the states in vicinity of Fermi level (Fig. 4). The sharp peak is related with the single $2p^1$ electron localized on non-oxidized boron atom surrounded by oxidized boron atoms. Survival of the magnetic moment are caused by the absence of the second boron atom suitable for the oxidation in vicinity of magnetic center. To evaluate magnetic interactions between remote magnetic centers we multiply the supercell along a-axis and calculate the difference between total energies of the systems with parallel and antiparallel orientation of the spins on magnetic centers. Calculated difference evidence that magnetic interactions between the centers is ferromagnetic. The magnitude of the energy difference between ferro- and antiferromagnetic configurations is 5 meV. Discussed magnetic moments in OBS can be eliminated by monovalent species. For the check of this possibility, we modeled hydrogenation of boron atom with magnetic moment in β-OBS (Fig. 3a). Saturation of dangling bond provides elimination of magnetic moment and appropriate features in electronic structure because now unpaired electron on *2p* shell participate in covalent bond with 1s electron of hydrogen that provides shift down of the corresponding energy level (Fig. 4). In result of the hydrogenation of magnetic center the value of calculated within GGA approach bandgap increased from about 0.1 to almost 0.4 eV. Note that DFT provides underestimation of the bandgap. To estimate the values of the real bandgap we will used relation between GGA and experimental values discussed in the Ref. [69] Thus the calculated value of 0.1 eV is corresponding with experimental value of about 0.3 eV and calculated 0.4 eV with the real bandgap of 0.8 eV. Note that discussed passivation is endothermic process and require the energy of +0.53 eV/H. Thus magnetism in OBS is chemically stable even at room temperature. This combination of air-stable ferromagnetism and narrow bandgap makes β-OBS prospective material for spintronic applications but further experimental exploration of the reproducibility of magnetism in OBS is required. In the case of non-magnetic χ-OBS the value of the calculated bandgap is larger (calculated value ~1 eV that is corresponding with the experimental values of order of 1.8 eV) which in combination of low weight and flexibility makes this material attractive as material for solar cells. The multiple peaks around Fermi energy (between -2 and +2 eV) could be the source of multiple optical transitions [64] that makes OBS optical-active material (see next subchapter). The appearance of these peaks is related with formation of localized states of the electrons on bonds between non-oxidized boron atoms (see Fig. 3 and inset Fig. 4). Partial oxidation provides break of the continuous web of the overlapping π-orbitals [39-42] of boron atoms that lead to the opening of the bandgap. Thus the number of the peaks near Fermi level and the distance between the peaks depends from initial atomic structure of borophene that guide oxidation process and amount of non-oxidized boron atoms and as results localized *$2p^1$* states near Fermi level.

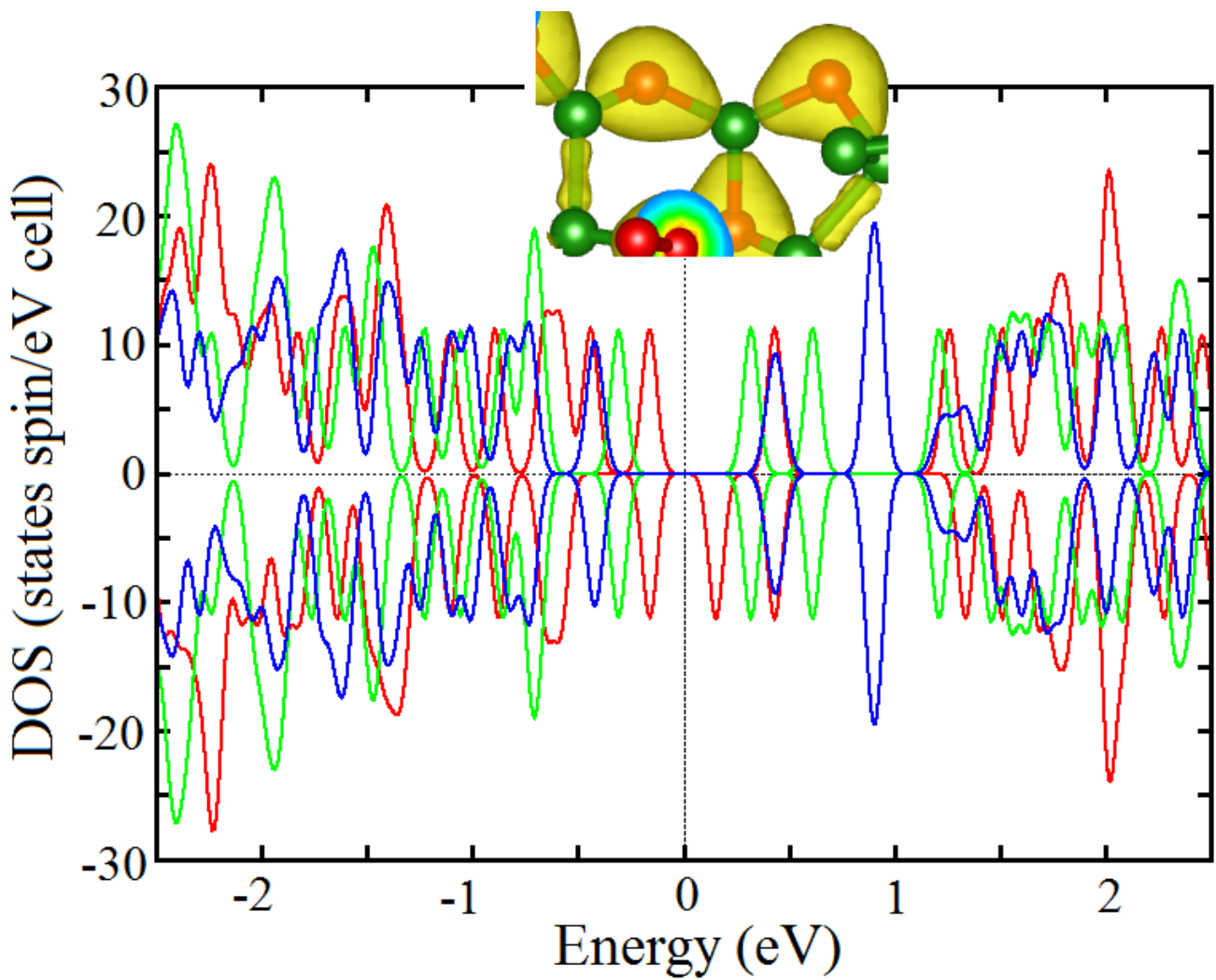

**Figure 4.** Spin-polarized density of states of fully oxidized boron sheet of β-type (red line) and χ-type (blue line). Green line is corresponding with β-OBS after passivation of magnetic center by hydrogen (Fig. 3a). Charge density distribution in the part of oxidized β-OBS are shown on inset.

**3.4. Adsorption of molecules on oxidized boron sheets**

The next step of our survey of the physical and chemical properties of OBS is the study of the conditions of adsorption of various molecules on both studied types of OBS and influence of the adsorption on electronic structure. In our modelling we considered three types of the adsorbants – nitrogen dioxide, water and hydrogen. First, we calculated enthalpy of adsorption by the same formula that was used for the calculation of the formation energy. Results of calculations (see Tab. I) demonstrate robust adsorption of nitrogen dioxide and water on β-OBS and less stable adsorption on χ-OBS.

**Table I.** The enthalpies of adsorption (eV/mol) of selected gases on OBSs.

| Adsorbant/Substrate | β-OBS  | χ-OBS  |
|---------------------|--------|--------|
| $NO_2$              | -0.357 | -0.170 |
| $H_2O$              | -0.386 | -0.119 |
| $H_2$               | -0.008 | -0.059 |

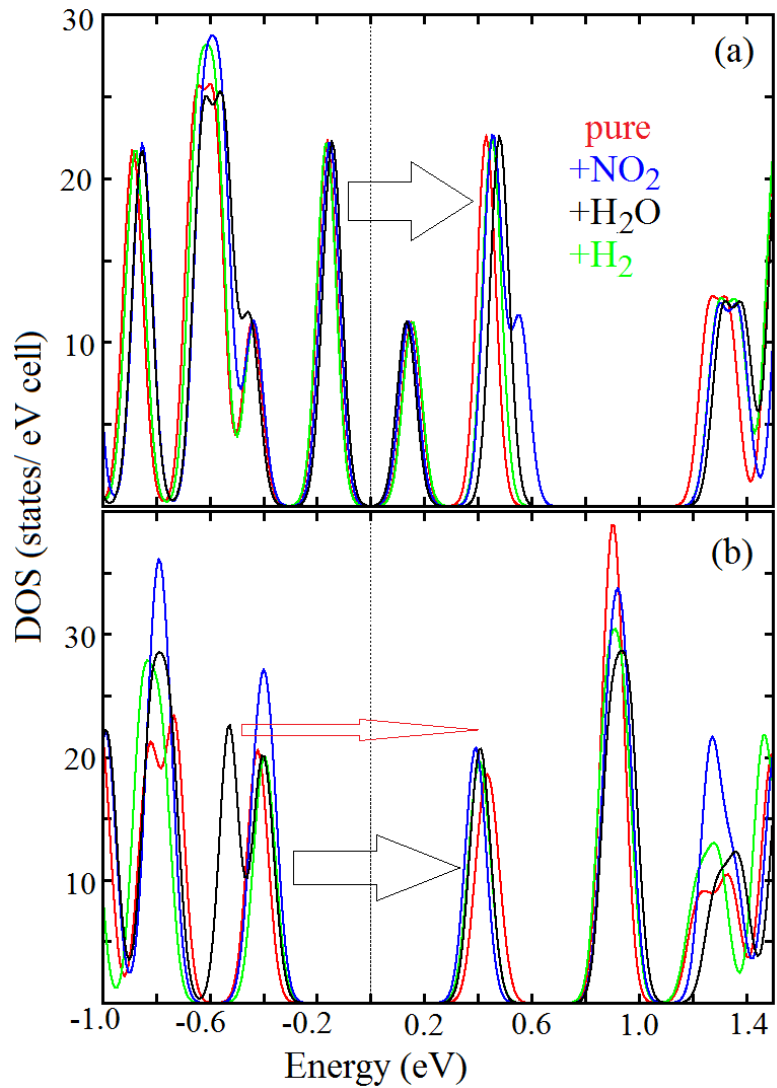

**Figure 5.** Total densities of states of β-OBS (a) and χ-OBS (b) before (red) and after adsorption of nitrogen dioxide (blue), water (black) and hydrogen (green) molecules. Arrows are indicating possible optical transitions influenced by adsorption of the molecules.

For the hydrogen molecule stable adsorption will occur only at χ-OBS but the magnitude of the enthalpy of the process is rather small. Unstable adsorption of hydrogen molecule on β-OBS provides no changes in electronic structure in vicinity of Fermi level. Contrary, all considered cases of the stable adsorptions provide visible changes in the positions and shapes of some peaks in vicinity of Fermi level that should change the intensity and perhaps quench some optical transitions (the key transitions are shown by arrows in Fig. 5). In the case of β-OBS will occur blue shift in adsorption and in the case of χ-OBS red shift. Note that similarly to other boron nanostructures [27,28] both studied OBS systems changes in the electronic structure is distinct for each type of adsorbed molecule therefore OBS can be proposed as sensing material with high selectivity. Adsorption of water provides appearance of the states at about -0.6 eV that can create additional transitions. So we can propose OBS as chemically stable humidity sensors.

### 3.6. Membrane properties

Combination of porous structure and chemical stability of OBSs creates opportunities for possible usage of these materials as membranes for the separation of the gases. For evaluation of permeability of OBSs we performed calculation of migration of hydrogen atom and hydrogen protons throughout the centers of the pores in both considered OBSs. For the simulation of the proton we generate pseudopotential for hydrogen atom with $0.05e^-$ on $1s$ orbital and perform the calculation of the system with the charge $+0.95e^-$. This method was previously used for the study of the migration of the proton trough graphene and hexagonal boron nitride. [70] Results of the calculations demonstrate energy barrier for $H^+$ is 0.704 eV for β-OBS and 0.594 eV for χ-OBS. For atomic hydrogens the values of the energy barriers are 0.751 and 0.693 eV. Thus OBSs could be proposed as impermeable material for hydrogen material and can be used for the coating of the volumes for hydrogen storage and flowing to prevent hydrogen leakage.

### 4. Conclusions

First principles modelling of the interaction of borophene sheets with molecular oxygen demonstrate instability of these materials in free standing form to oxidation. In result of the exothermic oxidation, borophene sheets will turn to foam-like boron oxide membrane with incorporated non-oxidized boron clusters. Obtained material is narrow-gap semiconductor and can be used as light-weight chemically stable solar cells. In some kinds of borophene sheets oxidation provide the appearance of stable at ambient condition *d0*-magnetism caused by unpaired electrons on single non-oxidized boron atoms. Stable physical adsorption of various gases on oxidized borophene sheets leads to the changes in optically active part of borophene spectra that make this material attractive for gas sensing. Calculated barrier of the migration of hydrogen protons and atomic hydrogen throughout the pores of oxidized borophene demonstrates impermeability of the membrane for both considered species that make possible usage of oxidized borophene as hydrogen-leakage preventing coating.